\documentclass[aps,prc,twocolumn,showpacs,superscriptaddress,nofootinbib,nolongbibliography]{revtex4-2}  % for review and submission
\usepackage{changes} 
\usepackage[para,online,flushleft]{threeparttable}
\usepackage{graphicx}  % needed for figures
\usepackage{dcolumn}   % needed for some tables
\usepackage{bm}        % for math
\usepackage{amssymb}   % for math
\usepackage{amssymb,amsmath,amsfonts} % for math
\usepackage[breaklinks=true,debug=true]{hyperref}
\usepackage{braket}    % for QM notation
\usepackage{multirow}
\usepackage{adjustbox}
\usepackage{gnuplottex}
\usepackage[T1]{fontenc}
\usepackage[utf8]{inputenc}
\usepackage{color}
\usepackage{dsfont}
\usepackage{tikz}
\usepackage{filecontents}
\usepackage{pgfplots}

\usepgfplotslibrary{groupplots}
\pgfplotsset{compat=newest}
\usepgfplotslibrary{units}

\begin{document}

%\widetext
%\leftline{Version 02 as of \today}
%\leftline{To be submitted to PRC}

\title{Improved determination of the oscillator parameters in nuclei}

\author{L.~Xayavong} 
\email{xayavong.latsamy@yonsei.ac.kr}
\affiliation{Department of Physics, Yonsei University, Seoul 03722, South Korea}
\author{Y.~Lim}
\email{ylim@yonsei.ac.kr}
\affiliation{Department of Physics, Yonsei University, Seoul 03722, South Korea}
%\affiliation{Physics Department, Faculty of Natural Sciences, National University of Laos, 7322 Dongdok, Vientiane Capital, Lao PDR}
%\author{N.~A.~Smirnova} 
%\email{smirnova@lp2ib.in2p3.fr}
%\affiliation{CENBG (CNRS/IN2P3 -- Université de Bordeaux), 33175 Gradignan cedex, France}
\vskip 0.25cm  
\date{\today}

\begin{abstract} 

The oscillator parameter in nuclei is refitted to reproduce the available charge radius data. As an important improvement, we include the Coulomb term evaluated within the assumption of a uniformly charged sphere, and take into account the symmetry effect induced by the difference between $N$ and $Z$ numbers in a straightforward manner using conventional parameterization. The Coulomb interaction has repulsive effect, causing the wave functions to extend further toward the nucleus exterior, resulting in an effectively larger oscillator length parameter. The symmetry effect is attractive for protons in neutron-rich nuclei and for neutrons in proton-rich nuclei, and repulsive for the other cases. Therefore, three distinct oscillator parameters are determined: one for protons, one for neutrons, and one isospin-invariant version, which is obtained by subtracting the Coulomb and symmetry contributions. Additionally, we explore the direct fit of the harmonic oscillator wave functions to the eigenfunctions of the Hartree-Fock mean field using the Skyrme interaction. Generally, this method agrees well with the others for light nuclei, typically up to $^{40}$Ca. Beyond this nucleus, however, the results begin to diverge over the orbits chosen for the fit. Only the parameters values obtained for the last occupied states agree remarkably well with the conventional ones throughout the mass range under consideration.

\end{abstract}

%\pacs{21.60.Cs, 23.40.Bw, 23.40Hc, 27.30.+t}
\maketitle

\section{Introduction}

The oscillator potential itself is not very realistic in nuclear physics. However, it provides an analytical solution to the one-body Schr\"{o}dinger equation and retains all symmetries of the atomic nucleus. Therefore, it is often a preferable choice for generating the single-particle basis for solving nuclear many-body problems, especially within the nuclear shell model and other variants of the configuration interaction theory. The oscillator potential is characterized by the angular frequency $\omega$ which can be converted into the length parameter $b$ using the relationship $b=\sqrt{\hbar/m\omega}$ where $\hbar$ is the reduced Planck's constant and $m$ is the nucleon's mass. The full harmonic-oscillator Hamiltonian for a particle without spin reads 
\begin{equation}\label{ho}
H_{HO} = \frac{\bm{p}^2}{2m} + \frac{1}{2}m\omega^2\bm{r}^2, % - \zeta \bm{l}\cdot\bm{s} - \kappa \bm{l}^2, 
\end{equation}
with the momentum operator defining as $\bm{p}=-i\hbar\bm{\nabla}$. 

In principle, nuclear structure calculations using a microscopic many-body method should be independent on the choice of basis functions if the configuration space is sufficiently large (ideally, the full Hilbert space). However, it is always difficult for this condition to be fulfilled due to computational limitations. 
For example, in the shell model which employs a full configuration space, 
%the $M$-scheme, 
%only few nucleons outside doubly closed-shell core can be treated as active 
%because the size of configuration space 
basis dimensions increase almost exponentially with a nucleon number. 
Because of this reason, except for very light nuclei, calculations are performed only for valence nucleons in a model space typically consisting of one oscillator shell. Even such model spaces become prohibitive for nuclei with $A>100$.
In this situation, an accurate determination of potential parameters may help to considerably improve a nuclear model's predictions. 

The oscillator parameters have been chosen in accordance with global systematics of nuclear charge radii. 
A traditional and widely-used prescription is that of Blomqvist and Molinari~\cite{Blomqvist1968} where $b^2$ is expressed as a function of mass number ($A$):
\begin{equation}
b^2 = 0.90A^\frac{1}{3} + 0.70 ~\text{fm}^2, 
\end{equation}
while not taking into account the difference between proton and neutron numbers. 

A more refined prescription for $b^2$ has been established by Kirson~\cite{Kirson2007}. As an extension to the previous work, the author introduced five corrective terms into the mean squared radii of the point-like proton distribution (denoted as $\braket{\bm{r}^2_\pi}_{pt}$) before fitting to the measured mean squared charge radii,  
\begin{equation}\label{ch}
\braket{\bm{r}^2}_{ch} = \braket{\bm{r}^2_\pi}_{pt} - \frac{3b_\pi^2}{2A} + \braket{\bm{r}^2_\pi}_0 + \frac{N}{Z}\braket{\bm{r}^2_\nu}_0 + \frac{3\hbar^2}{4m^2c^2} + \Delta_{ls}, 
\end{equation}
where the subscripts $\pi(\nu)$ refer to proton(neutron), and the quantities $\braket{\bm{r}^2_\pi}_0(\braket{\bm{r}^2_\nu}_0)$ are the mean squared radii of a single proton(a single neutron). The term $-{3b_\pi^2}/{2A}$ is the correction due to the center-of-mass motion, whereas ${3\hbar^2}/{4m^2c^2}$ and $\Delta_{ls}$ are, respectively, the Darwin-Foldy and the relativistic spin-orbit contributions. His analysis yields the expression
\begin{equation}
b_\pi^2 = 0.983(4)A^\frac{1}{3} + 0.373(23)~\text{fm}^2, 
\end{equation}
for protons, and 
\begin{equation}\label{nn}
b_\nu^2 = 0.859(5)A^\frac{1}{3} + 0.699(24)~\text{fm}^2, 
\end{equation}
for neutrons. The $b_\nu^2$ expression was determined via the introduction of neutron skin thickness for $N\ne Z$ nuclei into Eq.~\eqref{ch}. 
See Ref.~\cite{Kirson2007} for more details. 

The purpose of this paper is threefold. Firstly, to update the experimental data on charge radii to be used in Eq.~\eqref{ch}. Secondly, to investigate the impact of the Coulomb repulsion, as well as the difference between neutron and proton numbers, on $\braket{\bm{r}^2_\pi}_{pt}$ and the subsequent oscillator parameter using an exact treatment. Thirdly, to explore an alternative method for fitting the length parameter, namely by maximizing the overlap integral between harmonic oscillator and realistic Skyrme-Hartree-Fock (SHF) radial wave functions. Our detailed methodology and discussions of the results are given in Section~\ref{method}. We present our conclusion and perspective in Section~\ref{conc}. 
% of neutron $s_{1/2}$ orbitals, for which Coulomb and spin-orbit terms are not present. 

\section{Methods for fitting the oscillator length parameter}\label{method} 
\subsection{Method I}\label{sub1}

Our first method follows the conventional framework~\cite{Kirson2007,Blomqvist1968}, which employs experimental data on charge radii to constrain the oscillator parameter, while including the five corrective terms discussed in the previous section. The main contributor, $\braket{\bm{r}^2_\pi}_{pt}$ in Eq.~\eqref{ch} is evaluated using harmonic oscillator radial wave functions, namely 
\begin{equation}\label{r}
\braket{\bm{r}^2_\pi}_{pt} = \displaystyle \frac{1}{Z} \sum_{nl} N_{nl}^\pi \braket{\psi_{nl}^\pi|\bm{r}^2|\psi_{nl}^\pi}, 
\end{equation}
where the sum is taken over all occupied states of protons. 
%\added{The index $\pi$ is added to the wave functions to distinguish between protons ($\pi$) and neutrons ($\nu$)}. 
The single-particle matrix element, 
\begin{equation}\label{ex1}
\braket{\psi_{nl}^\pi|\bm{r}^2|\psi_{nl}^\pi}=(2n+l+\frac{3}{2})b_\pi^2, 
\end{equation} 
has been derived using the well-known virial theorem with $n$ and $l$ denoting the radial and orbital angular momentum quantum numbers, respectively. The proton occupation numbers $N_{nl}^\pi$ are fixed within the so-called equal-filling approximation. 
%The impact of the occupation of intruder level which is pushed down due to spin-orbit coupling is found to be insignificant and is neglected for the present study. More details on this effect were discussed in Ref.~\cite{Kirson2007}. 
%% 
%We add the superscript `$I$' to the oscillator length parameter to indicate the fitting method. 
Substituting these expressions into Eq.~\eqref{ch}, we obtain
\begin{equation}\label{bI}
b_\pi^2 = \frac{1}{f}\frac{Z}{\tilde{Z}} \left( \braket{\bm{r}^2}_{ch} - \braket{\bm{r}^2_\pi}_0 - \frac{N}{Z}\braket{\bm{r}^2_\nu}_0 - \frac{3\hbar^2}{4m^2c^2}  \right), 
\end{equation}
where $\tilde{Z}=\sum_{nl} N_{nl}^\pi(2n+l+3/2)$ and $Z/\tilde{Z}\le 1$. The factor $f=[1-3Z/(2A\tilde{Z})]$ accounts for the center of mass motion and has an effect of enlarging $b_\pi^2$ especially in the light mass region. The relativistic spin-orbit contribution ($\Delta_{ls}$) is neglected for simplicity. 

\begin{figure*}[ht!]
 \centering
\includegraphics[width=\textwidth]{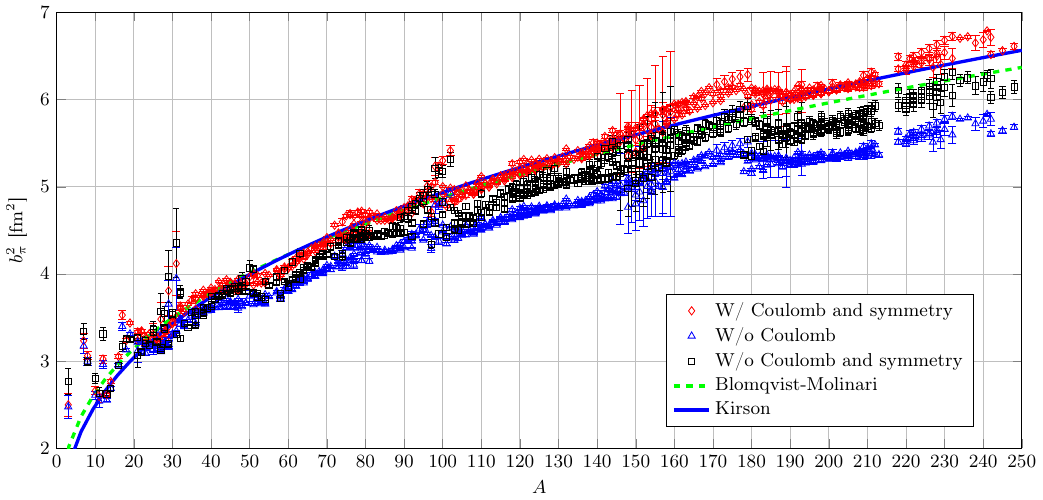}
\caption{\label{fig1}(Color online) Illustration of the Coulomb and symmetry effects on the oscillator length parameter for protons. The $b_\pi$ values obtained with Method I correspond to those marked with the diamond symbol. The results from Blomqvist-Molinari~\cite{Blomqvist1968} and Kirson~\cite{Kirson2007} are also given for comparison.}
\end{figure*}

The experimental 
charge radius data for 797 nuclei from $A=2$ (excluding $^1$H) to $248$ have been taken from the latest compilation of Angeli and Marinova~\cite{AngMari2013}. The updated data in Refs.~\cite{Li2021,Ohayon2022} are also considered. 
Our results for $b_\pi^2$ agree very well with those of Kirson as one can see from Fig.~\ref{fig1}. 
It should be noted, however, that our calculation in this section includes both open-shell and closed-shell nuclei, therefore the obtained $b_\pi^2$ values are significantly scattered off their trend line in the regions where deformation and correlation are dominant. 
%The fitted errors of our result are shown on the right panel of Fig.~\ref{fig1}. 
On the other hand, the large scattered points near the coordinate origin %of the plot (left panel) 
might be due to the breakdown of the mean field theory towards the $A\to 1$ limit. 
It is also seen that our calculation slightly underestimates the result of Blomqvist and Molinari~\cite{Blomqvist1968} in the light-mass region,
but overestimates it for nuclei with a mass number starting around $A=70$. 
%%
%%
% (the trendency of our result is almost coincided with green curve in Fig.~\ref{fig1}). 
Within this method, we obtain the following expression for protons:
\begin{equation}\label{piI}
\begin{array}{ll}
%b_\pi^2 = 1.003(4) A^\frac{1}{3} + 0.275(19) ~\text{fm}^2.
b_\pi^2 &= \displaystyle 0.214(20) + 1.034(5) A^\frac{1}{3} \\[0.1in]
& \displaystyle- 1.554(144) I + 5.634(544) I^2~\text{fm}^2.
\end{array}
\end{equation}
with $\chi^2/\nu=48.869$. The number of degrees of freedom, $\nu$ is obtained by subtracting the number of model parameters (4 parameters) from the sample size (1068 nuclei). Beyond the conventional form, which includes only a constant and a term proportional to $A^\frac{1}{3}$, we introduce linear and quadratic terms in $I$ to distinguish between isobars where $I=(N-Z)/A$. Without this extension, our fit would yield $\chi^2/\nu=54.197$. 
A large value of $\chi^2/\nu$ ($\chi^2/\nu\gg 1$) indicates that the uncertainties in the data sample are effectively smaller than the distances of individual data points from their trend line. %\added{Could you please clarify `tendency'? I'm not sure about `tendency'. Is it related to a fitting function? }
To account for this inconsistency, we scale the obtained uncertainties on the model parameters with $\sqrt{\chi^2/\nu}$ as suggested by the Particle Data Group~\cite{10.1093/ptep/ptac097}. The small discrepancy between our results and 
those of Kirson arise mainly from the difference in the determination of $N_{nl}^\pi$ (see Ref.~\cite{Kirson2007} for details). 

In order to extract the oscillator length parameter for neutrons, we follow Kirson~\cite{Kirson2007}, using the relation 
\begin{equation}\label{skin}
    \Delta r_{np} = \braket{\bm{r}^2_\nu}_{pt}^\frac{1}{2} - \braket{\bm{r}^2_\pi}_{pt}^\frac{1}{2}, 
\end{equation}
where $\Delta r_{np}$ is the neutron skin thickness. The squared radius of point-like neutron distribution $\braket{\bm{r}^2_\pi}_{pt}$ can be expressed as Eq.~\eqref{r} and Eq.~\eqref{ex1} for protons except that the proton occupation number $N_{nl}^\pi$ and the parameter $b_\pi$ must be replaced with those of neutrons. Inserting the expression of $\braket{\bm{r}^2_\nu}_{pt}$ and of $\braket{\bm{r}^2_\pi}_{pt}$ into Eq.~\eqref{skin}, we obtain 
\begin{equation}\label{nI}
    b_\nu = \left(\frac{N}{\tilde{N}}\right)^\frac{1}{2} \times \left[ \displaystyle \Delta r_{np} + b_\pi \left( \frac{\tilde{Z}}{Z} \right)^\frac{1}{2} \right]
\end{equation}
where $\tilde{N} = \sum_{nl} N_{nl}^\nu (2n + l + 3/2)$ with the sum running over all occupied states of neutrons. Therefore $N/\tilde{N}\le 1$. 
The neutron occupation number $N_{nl}^\nu$ is determined with the same method as $N_{nl}^\pi$. 
The first term on the right-hand-side (r.h.s) of Eq.~\eqref{nI} is induced by the neutron skin thickness whereas the following term is influenced by the difference between $N$ and $Z$. 
Unlike the work of Kirson which employed the empirical formula of $\Delta r_{np}$ extracted from antiproton interaction with nuclei~\cite{TRZCINSKA2004157} or hadronic atom and hadron scattering data~\cite{FRIEDMAN2005283}, we calculate this quantity within the Hartree-Fock-Bogoliubov method using effective Skyrme interaction. We include all even-even nuclei whose charge radius data are available from the above-mentioned compilations. A least squares fit to the results of these calculations yields 
\begin{equation}\label{np}
\begin{array}{ll}
%\Delta r_{np} = 1.094(24) \frac{(N-Z)}{A} - 0.059(4)~\text{fm}. 
\Delta r_{np} & \displaystyle= -0.032(2) + 0.808(41) I \\[0.1in]
& \displaystyle + 0.598(207) I^2 \pm 0.05_L~\text{fm} 
\end{array}
\end{equation}
with $\chi^2/\nu=1.616$. Note that we also exclude cases with $A<10$ from our mean field calculations; consequently, our sample size for fitting Eq.~\eqref{np} is reduced to 317. Then, the number of degrees of freedom in this process becomes $317-3=314$. This small $\chi^2/\nu$ value indicates that the model's errors are consistent with the uncertainties in the data sample. 
Several well-established Skyrme force parameterizations are considered, 
%\deleted{in these calculations}, 
%\deleted{namely SLY4 andSLY5~\cite{CHABANAT1998231}, SKM*~\cite{skm*}, SGII~\cite{VANGIAI1981379}, SII to SIV~\cite{BEINER197529}, and UNEDF0 to UNEDF2~\cite{PhysRevC.82.024313,PhysRevC.85.024304,PhysRevC.89.054314}.}
namely SLY4/SLY5~\cite{CHABANAT1998231}, SKM*~\cite{skm*}, SGII~\cite{VANGIAI1981379}, SII/SIII/SIV~\cite{BEINER197529}, and UNEDF0/UNEDF1/UNEDF2~\cite{PhysRevC.82.024313,PhysRevC.85.024304,PhysRevC.89.054314}.
%\added{Is that UNEDF2 or UNEDF1?}
Besides the spread of the results among the selected Skyrme parameterizations, we account for an uncertainty of $\pm 0.05$~fm for $\Delta r_{np}$ 
propagated from 
the symmetry energy slope ($L$) for $^{208}$Pb, which was estimated to be $64\pm39$~MeV~\cite{PhysRevLett.106.252501}. 
The pairing correlation and deformation are found to be significant for certain individual cases; however, their impact on the global trend of neutron skin thickness is generally negligible. 
By substituting the expression~\eqref{np} into Eq.~\eqref{nI} and then performing a least squares fit with a model similar to Eq.~\eqref{piI} for $b_\pi^2$, we obtain
\begin{equation}
\begin{array}{ll}
%   b_\nu^2 = 0.875(5) A^\frac{1}{3} + 0.653(23)~\text{fm}^2. 
b_\nu^2 & = \displaystyle 0.626(19) + 0.903(4) A^\frac{1}{3} -1.274(119) I \\[0.1in]
& \displaystyle+ 2.679(442) I^2~\text{fm}^2. 
\end{array}
\end{equation}
The resulting $\chi^2/\nu$ value is 0.861. We notice a somewhat diminished uncertainty in $b_\nu^2$ compared to $b_\pi^2$. This reduction is attributed to the factors $N/\tilde{N}$ and $(N/\tilde{N})\times (\tilde{Z}/Z)$ in Eq.~\eqref{nI}. Given that $N/\tilde{N}<1$ and $Z/\tilde{Z}<1$ as noted above, the factor $(N/\tilde{N})\times (\tilde{Z}/Z)$ is also effectively less than 1, owing to the predominance of neutron-rich species among the majority of nuclei. 
%\deleted{We notice that a large $\chi^2_\nu$ value is obtained for all linear fits carried out in this work (usually $\chi^2_\nu\ge 200$). This indicates that the estimated uncertainties are effectively much smaller than the distance from the individual data points relative to their trend line. To overcome this difficulty, we scale the obtained uncertainties on the model parameters with $\sqrt{\chi^2_\nu}$ as suggested by the Particle Data Group~\cite{10.1093/ptep/ptac097}}. 
Again, our result for $b_\nu^2$ is not very far from that of Kirson in Eq.~\eqref{nn}, even though we use a different method for the determination of neutron skin thicknesses. 

\subsection{Method II}\label{sub2}

Despite the unrealistic nature of the oscillator potential itself, the single-particle Hamiltonian~\eqref{ho} does not account for the repulsive Coulomb force among protons and the symmetry effect induced by the difference between neutron and proton numbers. This means that replacing the $\bm{r}^2$ operator in Eq.~\eqref{r} by unity, we will get a probability density distribution of uncharged particles of an $N=Z$ nucleus. 
To go beyond this conventional picture, we add the following Coulomb term derived from an assumption of a uniformly charged sphere, 
to the oscillator Hamiltonian: 
\begin{equation}\label{vc}
V_C(\bm{r}) = \frac{Ze^2}{2R_C}\left( 3 - \frac{\bm{r}^2}{R_C^2} \right) \left( \frac{1}{2} - t_z \right) , 
\end{equation}
where $e$ is the elementary charge and $t_z$ is the isospin projection of the nucleon with the convention of $t_z=\frac{1}{2}$ for neutrons and $-\frac{1}{2}$ for protons. The following convention for isospin must be applied when $t_z$ is used as an index: $t_z=\pi$ (for protons) or $\nu$ (for neutrons). In fact, Eq.~\eqref{vc} is correct only for $r<R_C$. 
%However, this does not matter since the potential $m\omega^2\bm{r}^2/2$ diverges very quickly to large distances.
However, the significance of this is negligible because 
the potential, $m\omega^2\bm{r}^2/2$ diverges rapidly as distances increase. 
For simplicity, the Coulomb exchange term is neglected. The Coulomb radius, $R_C$ is parameterized in literature as $R_C\approx 1.26~\text{fm}\times A^\frac{1}{3}$~\cite{SWV}. For the present work, we fix this parameter with the measured mean squared charge radii through the following formula~\cite{Elton}, 
\begin{equation}
R_C^2 = \frac{5}{3} \braket{\bm{r}^2_\pi}_{pt}, 
\end{equation}
where $\braket{\bm{r}^2_\pi}_{pt}$ can be converted into $\braket{\bm{r}^2}_{ch}$ using the relation Eq.~\eqref{ch}. 

\begin{figure*}
%   \begin{minipage}[t]{0.48\textwidth}
%     \includegraphics[width=\textwidth]{harmonic_oscillator_v3-figurec.pdf}
% %    \caption{}
%     \label{fig:}
%   \end{minipage}
%   \hfill
%   \begin{minipage}[t]{0.48\textwidth}
%     \includegraphics[width=\textwidth]{harmonic_oscillator_v3-figured.pdf}
% %    \caption{xxx}
%     \label{fig:}
%   \end{minipage}
%   \begin{minipage}[t]{0.48\textwidth}
%     \includegraphics[width=\textwidth]{harmonic_oscillator_v3-figurea.pdf}
% %    \caption{}
%     \label{fig:}
%   \end{minipage}
%   \hfill
%   \begin{minipage}[t]{0.48\textwidth}
%     \includegraphics[width=\textwidth]{harmonic_oscillator_v3-figureb.pdf}
% %    \caption{}
%     \label{fig:}
%   \end{minipage}
\centering
\includegraphics[width=\textwidth]{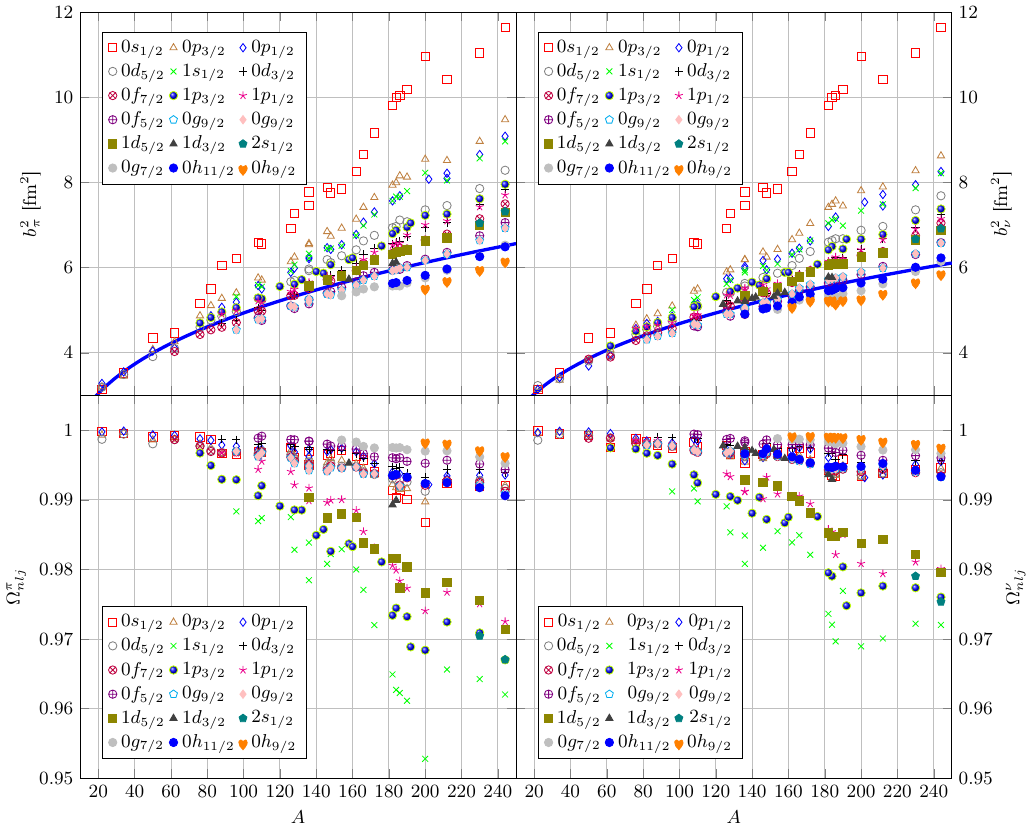}
\caption{\label{fig2}(Color online) Result of Method III for all occupied orbits of both closed-shell and open-shell nuclei. The fitted length parameter values are shown in the top-left (protons) and top-right (neutrons) panels. 
%, differentiated by colors according to single-particle states (blue representing lowest energy, progressing to violet for the highest energy). 
The corresponding values of the overlap integral are given in the bottom-left and bottom-right panels, respectively. The solid curves in the top-row panels represent the results of Method I, neglecting all $(N-Z)/A$-dependent terms. 
%\added{Can you change the symbols for some of them?
%Everything is denoted by a squares. It may look better to use square, `x`, and triangle.} 
}
\end{figure*}

The first term on the r.h.s. of Eq.~\eqref{vc} does not depend on $\bm{r}$, so it has no effect on the eigenfunctions. Meanwhile, the second term is proportional to $\bm{r}^2$, so it can be merged with the oscillator potential. Similarly, the symmetry effect in an $N\ne Z$ system can be accounted for by introducing a factor $[1-2t_z\kappa(N-Z)/A]$ to the oscillator potential, such that 
\begin{equation}
\frac{1}{2}m\omega^2 \bm{r}^2 \to \frac{1}{2}m\omega^2\left[1-2t_z\kappa\frac{(N-Z)}{A}\right] \bm{r}^2
\end{equation}
where $\kappa$ is a free parameter characterizing the strength of the symmetry term. %The upper and lower signs correspond to proton and neutron, respectively. 
In principle, the spin-orbit term of Thomas~\cite{1926Natur.117..514T} can also be added in a straightforward fashion. However, within the oscillator potential the radial form of this spin-orbit term is reduced to a constant, so it has no effect on the radial component of wave functions, regardless of energy splits between spin-up and spin-down states. 

Therefore, the inclusion of the Coulomb and symmetry terms simply results in an effective oscillator frequency, namely
\begin{equation}\label{eff}
\omega_{t_z}^2 = \omega^2\left[ 1-2t_z\kappa\frac{(N-Z)}{A} \right] - \left( \frac{1}{2} - t_z \right) \frac{Ze^2}{mR_C^3} , 
\end{equation}
with $b_{t_z}^2 = \hbar/(m\omega_{t_z})$ and $b^2 = \hbar/(m\omega)$. 
%The parameter $\delta_\pm = \frac{1}{2} \pm \frac{1}{2}$ switches on the Coulomb repulsion for protons and switches off for neutrons. 
Since the Coulomb and symmetry terms are separated out, the parameter $b$ or $\omega$ in Eq.~\eqref{eff} should be isospin-invariant. 
Subsequently, the following relation can be derived: 
\begin{equation}\label{pm}
\omega_\pi^2 = \omega_\nu^2\frac{\displaystyle\left[ 1+\kappa\frac{(N-Z)}{A} \right]}{\displaystyle\left[ 1-\kappa\frac{(N-Z)}{A} \right]} - \frac{Ze^2}{mR_C^3} .
\end{equation}
These effective oscillator parameters correspond to those extracted from the experimental data within the method I in the previous subsection. Therefore, with a known $b_\pi$ and $b_\nu$, Eq.~\eqref{pm} may be regarded as an alternative tool for the determination of the symmetry parameter $\kappa$. 
Nevertheless, we found that the $\kappa$ values extracted from Eq.~\eqref{pm}, i.e. $\kappa=0.423(441)$, vary strongly from nucleus to nucleus and their average underestimates the values used in literature considerably. 
For the present calculations, we adopt $\kappa=0.75(12)$ which is taken from the global fits of the Woods-Saxon parameter sets~\cite{SWV,PhysRevC.97.024324,PhysRevC.26.1712,ROST1968184,Dudek_1979}. 
The resulting isospin-invariant $b^2$ values are represented by the expression 
\begin{equation}\label{iso}
\begin{array}{ll}
%b^2 = 0.857(5) A^\frac{1}{3} + 0.693(17)~\text{fm}^2.
b^2&\displaystyle=0.846(12)+0.760(5)A^\frac{1}{3}+1.688(236)I \\[0.1in]
&\displaystyle+12.759(1060) I^2~\text{fm}^2, 
\end{array}
\end{equation}
with $\chi^2/\nu=6.853$ and $\nu=1064$. 
Note that $b$ would correspond to the length parameter obtained for neutrons in subsection~\ref{sub1} if only nuclei with $N=Z$ (excluding the symmetry effect) are considered. Additionally, within Eq.~\eqref{pm}, $b^2$ can be extracted from either $b^2_\pi$ or $b^2_\nu$. For this study, we adopt the average of these two values and treat their difference as an uncertainty source. 

As a general feature, the inclusion of the Coulomb term leads to a reduction in the effective oscillator frequency or an increase in the effective length parameter, because it reduces the proton binding energies. 
%\deleted{The symmetry term has an opposite effect for protons in neutron-rich nuclei and its strength is found to be about 50~\% of the Coulomb strength.} 
The symmetry term has an opposite effect on protons in neutron-rich nuclei, 
contributing about 50~\% to the squared length parameter, as depicted in Fig.~\ref{fig1}, compared to the Coulomb contribution.
%\added{How can you quantify 50\% of the Coulomb strength?}
Consequently, a substantial cancellation occurs between the Coulomb and symmetry contributions almost everywhere throughout the mass range under consideration. 
The symmetry effect on neutrons in neutron-rich nuclei is repulsive, thus leading to a larger effective length parameter for neutrons. The impact of the Coulomb and symmetry terms on the effective length parameter for protons is illustrated in Fig.~\ref{fig1}. 

These results suggest that a distinction of the effective oscillator parameters between protons and neutrons, $b_\pi \ne b_\nu$, should be made for a calculation in which isospin-symmetry breaking is taken into account. This distinction would be particularly important for the shell model description of isospin mixing, where the configuration space is extremely limited. Conversely, a single oscillator parameter, $b$ should be used instead when isospin symmetry is assumed, such as in the conventional shell model calculations~\cite{usdab,gx1a,jun45}. 
%\added{A typical application of our results is given in Section xx}.

\begin{table}[ht!]
\setlength{\extrarowheight}{0.08cm}
\caption{\label{tab1} Results of the method III for $s_{1/2}$ states of closed-shell and closed-subshell nuclei. $E_{nlj}^\nu$ and $\Omega_{nlj}^\nu$ denote, respectively, the SHF single-particle energies of neutrons and the overlap integral between the SHF and oscillator wave functions, both averaged over all selected Skyrme parameterizations. The unit of the length parameter values is fm$^2$. The energies are in MeV. The roman numbers (I) and (III) indicate the methods of evaluations. The Coulomb repulsion is excluded in these calculations, resulting in $E_{nlj}^\nu=E_{nlj}^\pi$ and $b_\nu$ (III)=$b_\pi$ (III) in self-conjugate $N=Z$ nuclei. Consequently, the equality, $b_\nu$ (III)=$b_\nu$ (I) is expected for this comparison}.
\begin{threeparttable}
\begin{ruledtabular}
\begin{tabular}{cccccc}
Nuclei	&	states	&	$b_\nu^2$ (III)	&	$\Omega_{nlj}^\nu$	&	$E_{nlj}^\nu$	&	$b_\nu^2$ (I)	\\
\hline
$^{12}$C	&	$0s_{1/2}$	&	2.51(15)	&	0.99987	&	-36.822	&	2.693(22)	\\
$^{16}$O	&	$0s_{1/2}$	&	2.98(8)	&	0.99985	&	-37.542	&	2.901(22)	\\
$^{28}$Si	&	$0s_{1/2}$	&	3.18(14)	&	0.99993	&	-48.041	&	3.368(23)	\\
$^{32}$S	&	$0s_{1/2}$	&	3.63(30)	&	0.99949	&	-50.095	&	3.493(23)	\\
		&	$1s_{1/2}$	     &	3.54(30)	&	0.99696	&	-14.023	&			\\
$^{40}$Ca	&	$0s_{1/2}$	&	4.02(20)	&	0.99936	&	-49.739	&	3.714(23)	\\
		&	$1s_{1/2}$	&	3.78(20)	&	0.99852	&	-17.963	&			\\
$^{48}$Ca	&	$0s_{1/2}$	&	3.97(10)	&	0.99924	&	-50.823	&	3.770(33)	\\
		&	$1s_{1/2}$	&	3.86(10)	&	0.99895	&	-18.454	&			\\
$^{48}$Ni	&	$0s_{1/2}$	&	4.43(9)	&	0.99868	&	-54.904	&	4.194(33)		\\
		&	$1s_{1/2}$	&	3.94(9)	&	0.99641	&	-24.471	&			\\
$^{56}$Ni	&	$0s_{1/2}$	&	4.76(9)	&	0.99866	&	-55.451	&	4.081(24)	\\
	&	$1s_{1/2}$	&	4.02(9)	&	0.99744	&	-24.033	&		\\
$^{80}$Zr	&	$0s_{1/2}$	&	5.59(10)	&	0.99840	&	-56.431	&	4.517(26)	\\
	&	$1s_{1/2}$	&	4.66(10)	&	0.99451	&	-29.960	&		\\ 
$^{90}$Zr	&	$0s_{1/2}$	&	5.88(10)	&	0.99791	&	-56.803	&	4.564(30)	\\
	&	$1s_{1/2}$	&	4.74(10)	&	0.99294	&	-30.301	&		\\
$^{100}$Sn	&	$0s_{1/2}$	&	6.15(10)	&	0.99646	&	-59.503	&	4.817(27)	\\
	&	$1s_{1/2}$	&	4.98(10)	&	0.98870	&	-33.842	&		\\
\hline
$^{132}$Sn\tnote{$\dagger$}	&	$0s_{1/2}$	&	6.92(11)	&	0.99607	&	-57.472	&	5.072(48)	\\
	&	$1s_{1/2}$	&	5.54(11)	&	0.98565	&	-34.609	&		\\
$^{208}$Pb\tnote{$\dagger$}	&	$0s_{1/2}$	&	8.34(12)	&	0.99497	&	-60.193	&	5.827(44)	\\
	&	$2s_{1/2}$	&	6.02(12)	&	0.98544	&	-18.694	&		\\
$^{298}$Fi\tnote{$\dagger$}	&	$0s_{1/2}$	&	9.63(13)	&	0.99241	&	-60.671	&	6.506(50)	\\
	&	$1s_{1/2}$	&	8.42(13)	&	0.96707	&	-45.809	&		\\
	&	$2s_{1/2}$	&	7.06(13)	&	0.97486	&	-25.422	&		\\
$^{310}$Ubh\tnote{$\dagger$}	&	$0s_{1/2}$	&	9.78(13)	&	0.99513	&	-62.436	&	6.593(43)	\\
	&	$1s_{1/2}$	&	8.26(13)	&	0.97439	&	-47.760	&		\\
	&	$2s_{1/2}$	&	6.98(13)	&	0.979	&	-27.813	&		\\
\end{tabular}
\end{ruledtabular}
\begin{tablenotes}
{ \raggedright
\item[$\dagger$] Closed-shell or closed-subshell nuclei with $N\ne Z$}. 
\end{tablenotes}
\end{threeparttable}
\end{table}

\subsection{Method III}

One may further argue that, in addition to its unrealistic form in the space coordinate, the oscillator Hamiltonian~\eqref{ho} also misses an appropriate spin-orbit term. The presence of a spin-orbit term breaks the degeneracy of the oscillator states, leading to a difference in single-particle configurations, which then yields different expectation values for the mean squared radii. Furthermore, the experimental data of charge radii may contain a considerable contribution of deformation and other effects beyond the spherical mean field description. We therefore apply in this subsection an alternative method for determining the oscillator length parameter without the utilization of the experimental data. The idea is to directly fit the harmonic oscillator radial wave functions to the eigenfunctions of spherical SHF mean field by varying the length parameter while the Coulomb and nuclear charge-dependent forces are turned off. 
It is expected that the influence of the spin-orbit coupling, as well as spurious isospin mixing and beyond spherical mean field effects, can be avoided by considering only $s_{1/2}$ states of the doubly magic $N=Z$ nuclei, namely $^{16}$O, $^{40}$Ca, $^{56}$Ni and $^{100}$Sn. Although the sample size of this method would be too small for a prediction of the trend of the squared length parameter, it may provide a meaningful test of the previous methods through a point-by-point comparison. The fit of this method is carried out by maximizing the overlap integral defined below 
\begin{equation}\label{o}
\Omega_{nlj}^{t_z} = \int_0^\infty \psi_{nl}^{t_z}(r) \phi_{nlj}^{t_z}(r) r^2 dr, 
\end{equation}
where $\psi_{nl}^{t_z}(r)$ and $\phi_{nlj}^{t_z}(r)$ are, respectively, the harmonic oscillator and SHF radial wave functions. 
%\added{I'm wondering how you compare HO and SHF.
%HO obtained from the harmonic potential $1/2 kr^2$, right?
%What about SHF?}
%\added{==>Yes, HO wave function is eigenfunction of $1/2 kr^2$. SHF stands for 'Skyrme-Hartree-Fock'. More details on HF calculations are given in the following paragraph}
%%\deleted{Note that $\phi_{nlj}^{t_z}(r)$ is spin-independent} 
Note that the indices $l$ and $j$ in Eq.~\eqref{o} can be omitted
if only $s_{1/2}$ states are considered. Obviously, $\Omega_{nlj}^{t_z}$ reduces to the normalization integral if $\psi_{nl}(r)$ and $\phi_{nlj}^{t_z}(r)$ are perfectly coincided. 

For the SHF calculations, we employ the same sets of Skyrme parameterizations as considered in subsection~\ref{sub1}. 
Since the Coulomb repulsion is not included, the obtained oscillator length parameters must be compared with the isospin-invariant one given in Eq.~\eqref{iso} or the one obtained for neutrons within Method I. Note that these correspondences are valid only for self-conjugate nuclei, where the symmetry effect is absent. The results for the four above-mentioned doubly magic $N=Z$ nuclei are listed in Table~\ref{tab1}. For comparison, some other $N\ne Z$ species including closed-shell and closed-subshell nuclei are also listed. Unfortunately, it is found that the results are highly sensitive to the fine details of the SHF mean field due to the missing of the centrifugal barrier. 
Furthermore, the obtained length parameter values for a given doubly magic $N=Z$ nucleus vary remarkably depending on the radial quantum number, even when solely focusing on $s_{1/2}$ states. In principle, the state that generates the largest overlap integral, $\Omega_{nlj}$ should be selected, typically the $0s_{1/2}$ state, as evidenced by Table~\ref{tab1}. 
In particular, this method consistently yields $b$ values that tend to exceed those obtained with method I, except for light nuclei, typically below $^{40}$Ca. Generally, the largest $b$ value is obtained for the lowest state, specifically the $0s_{1/2}$, followed by a gradual decrease with increasing energy or radial quantum number.
This observation suggests that, considering specific higher single particle energy
occupied states with $l\ne0$ for particular nuclei could potentially enhance the agreement with method I, despite the lack of clear theoretical justification. In our opinion, the dependence on $n$, or more generally, on single-particle orbits, may be attributed to the weakly bound effect, which cannot be seen within the harmonic oscillator potential which rises to $\infty$ rapidly with increasing of the spatial distance. In addition, the eigenfunctions of $s_{1/2}$ states may still be influenced indirectly by the spin-orbit potential of the other occupied states with $l\ne 0$ due to the self-consistency of the SHF equation. 

In order to gain a better insight, we perform a further analysis with the inclusion of open-shell nuclei. We employ the standard SHF procedure with the presence of the Coulomb interaction, for which the Coulomb exchange contribution is evaluated using the Slater approximation. The obtained results are given in Fig.~\ref{fig2}. The $b_\nu$ values illustrated on the top-left panel of Fig.~\ref{fig2} are obtained by fitting to neutron eigenfunctions whereas the $b_\pi$ values on the top-right panel of the same figure are obtained by fitting to proton eigenfunctions. The corresponding values of the overlap integral are shown on the bottom panels of the same Figure. It is interesting to remark that the fitted values of $b_\nu$ and $b_\pi$ in Fig.~\ref{fig2} have a similar pattern except that $b_\pi$ is slightly larger than $b_\nu$ for the effects discussed in subsection~\ref{sub2}. In general, a lower SHF state yields a larger length parameter value, as noted for closed-shell nuclei in the above paragraph. We notice that a notable dependence on $l$ persists in the absence of the spin-orbit term, indicating a distinct impact of the centrifugal barrier on both the SHF and harmonic oscillator potential. Despite the dependence on $n$ and $l$, it is seen from Fig.~\ref{fig2} that only 
fitting to the last occupied states achieves reasonable agreement with the results obtained from method I for both neutrons and protons, irrespective of the weakly-bound effect. However, the corresponding overlap integral values do not consistently exhibit a similar pattern. On average, though, these values for the last occupied states tend to approach unity more closely than those for lower-energy states. With the inclusion of the Coulomb interaction and the consideration of opened-shell nuclei, the best-fit length parameter for last occupied states is represented by 
\begin{equation}\label{xx1}
\begin{array}{ll}
%\deleted{b_\pi^2 = 0.803 (13) A^\frac{1}{3} + 0.962(66)~\text{fm}^2},  
b_\pi^2 & \displaystyle= 0.355(74) + 1.007(18) A^\frac{1}{3} -2.581(652) I \\[0.1in]
&\displaystyle + 9.160(2234) I^2 ~\text{fm}^2, 
\end{array}
\end{equation} 
for protons, and by 
\begin{equation}\label{xx2}
\begin{array}{ll}
b_\nu^2 & \displaystyle= 1.097(76) + 0.773(21) A^\frac{1}{3} - 0.306(395)I \\[0.1in]
& \displaystyle + 2.466(1642)I^2 ~\text{fm}^2, 
\end{array}
\end{equation}
for neutrons. The corresponding $\chi^2/\nu$ values are 151.12 and 64.47, respectively. 
The isospin-invariant oscillator parameter can, in principle, be extracted from these results following the method described in subsection~\ref{sub2}. The resulting expression is, 
\begin{equation}\label{xx3}
\begin{array}{ll}
b^2 & \displaystyle= 1.201(48) + 0.705(13) A^\frac{1}{3} + 0.129(242)I \\[0.1in]
& \displaystyle + 3.994(1382)I^2 ~\text{fm}^2. 
\end{array}
\end{equation}
The parameterizations of the oscillator parameter determined by the present method differ considerably from those discussed in the previous subsections, in particular for neutrons, and the isospin-invariant version. 
The substantial uncertainties on the coefficients in Eq.~\eqref{xx1}, Eq.~\eqref{xx2}, and Eq.~\eqref{xx3} reflect a large scatter in the data samples. 
Nevertheless, these parameterizations are much improved compared to those obtained by fitting to $s_{1/2}$ states alone as listed in Table~\ref{tab1}. 
This improvement occurs despite the ambiguity in the fit quality, which is more pronounced around the Fermi level.

\section{Conclusion}\label{conc}  

In this study, we investigate three different methods for the evaluation of the oscillator length parameter. In the first method, we follow the conventional framework using the new updated data on charge radii as a constraint. The obtained result does not differ significantly from that of Kirson within the same method published in 2008. In the second method, we incorporate the Coulomb contribution based on the approximation of a uniformly charged sphere and the symmetry term contribution using existing parameterization. We found that the Coulomb repulsion has an effect of increasing the oscillator length parameter especially in heavy nuclei. Conversely, the inclusion of the symmetry term leads to an increase in the length parameter for protons in neutron-rich nuclei. Within the chosen parameterization, the symmetry contribution is found to be about 50~\% of the Coulomb contribution. In addition, the isospin-invariant length parameter has been appropriately extracted for the first time. 
In the last method, the oscillator length parameter is adjusted to maximize the overlap integral between the oscillator functions and the eigenfunctions of the SHF mean field. A remarkable agreement is obtained for light nuclei typically up to $^{40}$Ca. In heavier nuclei, the results are strongly orbit-dependent and only the fit of the last occupied states provides agreement with the conventional method. 

\begin{acknowledgments} 
We are extremely grateful to N. A. Smirnova for a careful reading of the manuscript. 
L. Xayavong and Y. Lim are supported by the National Research Foundation of Korea(NRF) grant funded by the Korea government(MSIT)(No. 2021R1A2C2094378). Y. Lim is also supported by the Yonsei University Research Fund of 2023-22-0126. 
%\added{N. A. Smirnova acknowledges the IN2P3/CNRS, France, in the framework of the ``Isospin-symmetry breaking'' and ``Exotic nuclei, fundamental interactions and astrophysics'' Master projects}. 
\end{acknowledgments}
%

%\bibliography{harmonic_oscillator}

\end{document}